\documentclass[showpacs,prl,aps]{revtex4}

\usepackage{amsmath}
\usepackage{amssymb}
\usepackage{latexsym}
\usepackage{amsfonts}
\usepackage{epsfig}
\usepackage{psfrag}
\usepackage{graphicx}
\usepackage{hyperref}

\usepackage{color}

\newcommand{\bea}{\begin{eqnarray}}
\newcommand{\ea}{\end{eqnarray}}
\newcommand{\eea}{\end{eqnarray}}

\begin{document}

\title{Schwinger Vacuum Pair Production in Chirped Laser Pulses}

\author{Cesim~K.~Dumlu}

\affiliation{Department of Physics, University of Connecticut,
Storrs CT 06269-3046, USA}

\begin{abstract}
The recent developments of high intensity ultra-short laser pulses have raised the hopes of observing Schwinger vacuum pair production which is one of the important non-perturbative phenomena in Quantum electrodynamics (QED). The quantitative analysis of realistic high intensity laser pulses is vital for understanding the effect of the field parameters on the momentum spectrum of the produced particles. In this study, we analyze chirped laser pulses with a sub-cycle structure, and investigate the  effects of the chirp parameter on the momentum spectrum of the produced particles. The combined effect of the chirp and carrier phase of the laser pulse is also analyzed. These effects are qualitatively explained by investigating the turning point structure of the potential within the framework of the complex WKB scattering approach to pair production.
\end{abstract}



\maketitle

\section{Introduction}

Schwinger vacuum pair production is one of the important predictions of QED which manifests itself as a non-perturbative effect of the external field applied to QED vacuum \cite{eh,schwinger}.  The probability of observing such a phenomenon becomes appreciable, when the external field's intensity reaches the scale of the critical field strength $E_{\rm cr}=\frac{m^2c^3}{e\hbar}\sim 10^{18}$ V/m. However, recent analyses suggest that vacuum pair production may be observed at much lower intensities by careful combining and shaping of the laser pulses \cite{Schutzhold:2008pz,Bell:2008zzb,Monin:2009aj,Heinzl:2010vg,DiPiazza:2009py,Bulanov:2010ei}. These analyses, together with plans for the experimental realization of field strengths one or two orders below the critical field strength \cite{tajima,gerstner}  have raised the hopes of observing such non-perturbative effect \cite{dunne-eli} . The observation of this effect is crucial for the understanding of other non-perturbative phenomena in quantum field theory such as cosmological particle production, \cite{parker} Unruh and Hawking radiation, \cite{ralf-eli,leonhardt} heavy ion collisions \cite{dima} as well as the Landau-Zener effect \cite{li} and non-perturbative effects in strong field AMO physics \cite{mourou}.

The quantitative analysis of Schwinger vacuum pair production, for fields varying in time, can be viewed as an over-the-barrier scattering problem in one dimension where the pair production rate corresponds to the reflection coefficient \cite{popov,breyzin}. This approach is mathematically equivalent to calculating the particle number expectation in a time dependent background field using the quantum Vlasov equation \cite{florian-1,kluger,schmidt1,schmidt2,Dumlu:2009rr}. In a recent study \cite{Hebenstreit:2009km}, using the latter approach, the momentum spectrum of the produced particles has been computed for an electric field of the form
\begin{equation}
E(t)=E_{0}\,e^{-\frac{t^{2}}{2\tau^{2}}}\cos\left(\omega t +\varphi\right)
\end{equation}
which includes a carrier phase $\varphi$ in addition to the frequency $\omega$ and the envelope pulse length $\tau$. The momentum spectrum was found to be extremely sensitive to these physical pulse parameters, and also to exhibit an interesting phase sensitivity between scalar and spinor QED \cite{Hebenstreit:2009km,florian-qfext09}.
This sensitivity can be understood in terms of quantum mechanical resonances, and a quantitative explanation in terms of complex WKB has recently been given in \cite{cg}.
Here we consider a more realistic laser field, incorporating the dependence on the chirp of the laser pulse, which effectively includes variations in the frequency, and we investigate the effect of such a chirp on the momentum spectrum. This is also important for the study of realistic fields from the experimental point of view; high intensity laser pulses are obtained through chirped pulse amplification technique \cite{Strick}, in which a short laser pulse is chirped and stretched by means of a dispersive stretcher, followed by the process of amplification and recompression of the pulse. During this process, there might be residual chirp in the field which may introduce frequency variations. Another motivation is due to interesting results of chirp effects in strong-field atomic and molecular physics \cite{weinacht,Krausz}.

\subsection{Chirped Electric Field}
\begin{figure}[htb]
\includegraphics[scale=0.55]{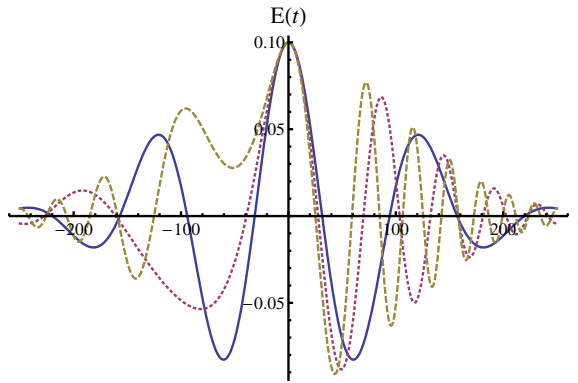}
\caption{The thick line shows the electric field (\ref{electric}) with chirp parameter $b=0$. The dotted [red] line has $b=0.00025$, and the dashed [gold] curve has  $b=0.0005$. All plots have $E_{0}=0.1$, $\omega=0.05$, $\tau=100$ (all the parameters are given in units of the electron mass $m$). Note that the shape of the pulse is very sensitive to the value of the chirp parameter  $b$.}
\label{efield}
\end{figure}

\begin{figure}[htb]
\includegraphics[scale=0.55]{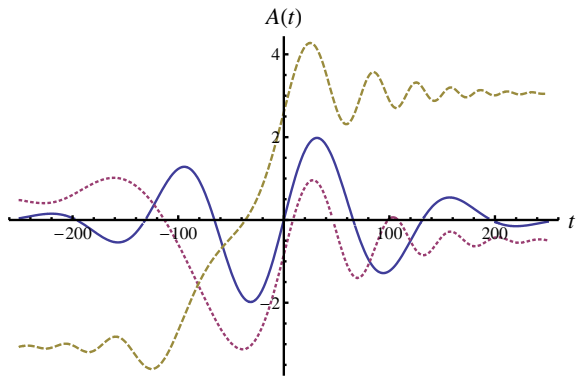}
\caption{The thick [blue] line shows the gauge field (\ref{gauge}) with chirp parameter $b=0$. The dotted [red] line has $b=0.00025$, and the dashed [gold] curve has  $b=0.0005$. All plots have $E_{0}=0.1$, $\omega=0.05$, $\tau=100$ . Note that the shape of the pulse is very sensitive to the value of the chirp parameter  $b$.}
\label{afield}
\end{figure}

We consider the electric field $\vec{E}(t)=(0, 0, E(t))$  as two counter propagating laser beams forming a standing wave and the possible effects due to spatial focussing are neglected. Here we analyze the field;
\begin{equation}
E(t)=E_{0}\, e^{-\frac{t^{2}}{2\tau^{2}}}\cos\left(b\,t^{2}+\omega\, t +\varphi\right)
\label{electric}
\end{equation}
where $b$ is the chirp parameter which causes a linear variation in the frequency. This type of field configuration is represented as $E(t)=-\dot{A}(t)$ where $A(t)$ is the gauge potential and given as:
\begin{eqnarray}
A(t)&=&-\frac{(-1)^{\frac{1}{4}}E_{0}\tau\sqrt{\pi/2}}{2(1+4b^{2}\tau^{4})}
e^{-\frac{i\omega^{2}\tau^{2}}{2i+4b\tau^{2}}}\left[f(t,\omega,\tau,\varphi,b)+g(t,\omega,\tau,\varphi,b)
\right]
\label{gauge}
\end{eqnarray}
where,
\begin{eqnarray}
f(t,\omega,\tau,\varphi,b) &=& e^{-\frac{2ib\omega^{2}\tau^{4}}{1+4b^{2}\tau^{2}}-i\varphi}
\sqrt{-i+ 2b\tau^{2}}(i+2b\tau^{2}) {\rm erfi}\left[\frac{\left(t(1+2i b\tau^{2})+\tau^{2}\omega\right)}
{\tau\sqrt{2-4ib\tau^{2}}}\right]\\\nonumber
g(t,\omega,\tau,\varphi,b) &=&e^{i\varphi}\sqrt{i + 2b\tau^{2}}(i+2b\tau^{2}) {\rm erfi}\left[\frac{\left(t(-i+2b\tau^{2})+\tau^{2}\omega\right)}
{\tau\sqrt{2-4ib\tau^{2}}}\right].
\end{eqnarray}
The imaginary error function is defined as:
\begin{equation}
{\rm erfi}(z) = -\frac{2i}{\sqrt{\pi}}\int^{iz}_{0}e^{-t^2}dt
\end{equation}
When the chirp parameter $b$ is set to zero, these fields reduce to those used in \cite{Hebenstreit:2009km}.
The effect of the chirp can be clearly seen in Fig \ref{efield} and \ref{afield}. The instantaneous frequency of the field linearly increases with the time for a constant chirp parameter $b$. The values of the other parameters such as field intensity $E_{0}$ and pulse width $\tau$ are chosen in the non-perturbative regime which may be realized experimentally and given in terms of normalized units ($\hbar=c=m=1$). Throughout this paper we use the values of $E_0$, $\omega$ and $\tau$ used in the previous work \cite{Hebenstreit:2009km,florian-qfext09}, adding various choices for the new chirp parameter $b$. So, we use $E_0=0.1 E_{cr}$, $\omega=0.05$ and $\tau=100$. All the field parameters are given in units of the electron mass $m$.
~

\subsection{Computational Formalism}
The calculation of the momentum spectrum of the particles produced by  a time dependent electric field can be done in several equivalent ways, which we summarize here. The fact that the field is assumed to be spatially uniform means that the spatial momentum $\vec{k}$ is a good quantum number and we can discuss each mode separately, as the production of a particle/anti-particle pair with momenta $\vec{k}$ and $-\vec{k}$.  For scalar QED, this involves a mode decomposition of the field operator as
\begin{eqnarray}
\Phi(\vec{x},t)=\int d^3k\,  e^{i\vec{k}\cdot \vec{x}}\left(u_{\bf{k}}(t)a_{\bf{k}}+u^{*}_{\bf{k}}(t)b^{\dagger}_{-\bf{k}}\right)
\label{mode}
\end{eqnarray}
Inserting this into the Klein-Gordon equation $(-D_{\mu}D^{\mu}+m^{2})\Phi=0$, yields the following "Schr\"odinger-like" equation for each mode coefficient function $u_{\vec{k}}(t)$:
\begin{eqnarray}
&\frac{d^2u_{\bf{k}}(t)}{dt^2} + \omega_{\bf{k}}^2(t)u_{\bf{k}}(t)=0
\label{schro}
\end{eqnarray}
where the effective time-dependent frequency of this oscillator problem is
\begin{eqnarray}
\omega_{\bf{k}}(t)=\sqrt{m^2+k_{\perp}^2+(k-\frac{q}{c}A(t))^{2}}
\label{omega}
\end{eqnarray}
Here $k$ and $k_\perp$ are the momenta along and perpendicular to  the direction of the electric field, respectively.
This Schr\"odinger-like problem has scattering boundary conditions, as follows from the usual Bogoliubov expansion \cite{breyzin,popov}. We make an adiabatic ansatz for (\ref{schro}) in terms of the basic leading WKB solutions as:
\begin{eqnarray}
u_{\bf{k}}(t)=\frac{\alpha_{\bf{k}}(t)}{\sqrt{2\omega_{\bf{k}}(t)}}
e^{-i\phi_{\bf{k}}(t)}+\frac{\beta_{\bf{k}}(t)}{\sqrt{2\omega_{\bf{k}}(t)}}e^{i\phi_{\bf{k}}(t)}
\end{eqnarray}
where
\begin{eqnarray}
\phi_{\bf{k}}(t) = \int_{-\infty}^{t}\omega_{\bf{k}}(t')dt'.
\label{phi}
\end{eqnarray}
We relate the two complex functions $\alpha_{\bf{k}}(t)$ and $\beta_{\bf{k}}(t)$ by requiring additionally that $\dot{u}_{\bf{k}}(t)$ follow the constant field form:
\begin{eqnarray}
\dot{u}_{\bf{k}}(t)=-i\omega_{\bf{k}}(t)
\left(\frac{\alpha_{\bf{k}}(t)}{\sqrt{2\omega_{\bf{k}}(t)}}e^{-i\phi_{\bf{k}}(t)}-\frac{\beta_{\bf{k}}(t)}{\sqrt{2\omega_{\bf{k}}(t)}}e^{i\phi_{\bf{k}}(t)}\right).
\end{eqnarray}
This requires that the Bogoliubov coefficient functions, $\alpha_{\bf{k}}(t)$ and $\beta_{\bf{k}}(t)$, satisfy the coupled equations
\begin{eqnarray}
\dot{\alpha}_{\bf k}(t)&=& \frac{\dot{\omega}_{\bf{k}}(t)}{2\omega_{\bf{k}}(t)}\beta_{\bf{k}}(t)e^{2i\phi_{\bf{k}}(t)}\\
\dot{\beta}_{\bf k}(t)&=& \frac{\dot{\omega}_{\bf{k}}(t)}{2\omega_{\bf{k}}(t)}\alpha_{\bf{k}}(t)e^{-2i\phi_{\bf{k}}(t)}
\label{coupled}
\end{eqnarray}
Note that these equations preserve the unitarity  condition
\begin{eqnarray}
|\alpha_{\bf{k}}(t)|^{2}-|\beta_{\bf{k}}(t)|^{2}=1
\label{unitarity}
\end{eqnarray}
which enforces the bosonic field commutation relations.
The number of particles produced in the mode ${\bf k}$ is given by
\begin{eqnarray}
{\mathcal N}({\bf k})=|\beta_{\bf k}(\infty)|^2
\label{n}
\end{eqnarray}
This particle number can be expressed in terms of the reflection coefficient for scattering in the Schr\"odinger-like problem (\ref{schro}):
\begin{eqnarray}
{\mathcal R}({\bf k})=\left| \frac{\beta_{\bf k}(\infty)}{\alpha_{\bf k}(\infty)}\right|^2=\frac{{\mathcal N}({\bf k})}{1+{\mathcal N}({\bf k})}\quad \Rightarrow \quad {\mathcal N}({\bf k})=\frac{{\mathcal R}({\bf k})}{1-{\mathcal R}({\bf k})}
\label{rn}
\end{eqnarray}
There are many equivalent ways to formulate this scattering computation. In the quantum kinetic equation (QKE) approach, one defines a "time-dependent particle number" $N({\bf k}, t)\equiv |\beta_{\bf k}(t)|^2$, whose time evolution is governed by the QKE, inherited from the coupled equations (\ref{coupled}). Alternatively, one can define a time-dependent reflection amplitude $R({\bf k}, t)=\beta_{\bf k}(t)/\alpha_{\bf k}(t)$, which satisfies the following Riccati equation  \cite{popov}, also inherited from the coupled equations (\ref{coupled}):
\begin{eqnarray}
\dot{R}_{\rm scalar}({\bf k}, t) = \frac{\dot{\omega}_{\bf k}(t)}{2\omega_{\bf k}(t)}\left(e^{-2i\phi_{\bf k}(t)} - R_{\rm scalar}^{2}({\bf k}, t)e^{2i\phi_{\bf k}(t)}\right),
\label{riccati-scalar}
\end{eqnarray}
where we have added the subscript ``scalar'' to distinguish from the spinor case below. Using the initial condition $R_{\rm scalar}({\bf k}, -\infty)=0$, one evolves (\ref{riccati-scalar})  to find ${\mathcal R}_{\rm scalar}({\bf k})\equiv | R_{\rm scalar}({\bf k}, \infty)|^2$, and hence ${\mathcal N}_{\rm scalar}({\bf k})$ from (\ref{rn}).

For spinor QED, an analogous construction exists, beginning with the Dirac equation instead of the Klein-Gordon equation, with the result again formulated either in terms of a QKE or a Riccati equation. Here we use the Riccati form, which for spinor QED reads \cite{Dumlu:2009rr,popov}:
\begin{eqnarray}
\dot{R}_{\rm spinor}({\bf k}, t) = -\frac{\dot{\omega}_{\bf k}(t)\,\sqrt{m^2+k_\perp^2}}{2\omega_{\bf k}(t)(k-A(t))}\left(e^{-2i\phi_{\bf k}(t)} + R^2_{\rm spinor}({\bf k}, t)e^{2i\phi_{\bf k}(t)}\right)
\label{riccati-spinor}
\end{eqnarray}
Here $\phi_{\bf k}(t)$ is the same function as was defined in (\ref{phi}).
Note the changes of sign with respect to the scalar QED Riccati equation (\ref{riccati-scalar}). This reflects the fact that in spinor QED the fields satisfy anti-commutation relations. Computationally, the consequence is that the relation between the particle number and the reflection coefficient becomes
\begin{equation}
{\mathcal N}_{\rm spinor}({\bf k})=\frac{\mathcal{R}_{\rm spinor}({\bf k})}{1+\mathcal{R}_{\rm spinor}({\bf k})},\hspace{0.2in}
\end{equation}
where ${\mathcal R}_{\rm spinor}({\bf k})\equiv | R_{\rm spinor}({\bf k}, \infty)|^2$, which is obtained by numerically integrating (\ref{riccati-spinor}).
In the subsequent sections, we study the particle number spectrum ${\mathcal N}({\bf k})$ for both scalar and spinor QED, obtained by integrating the Riccati equations (\ref{riccati-scalar}) and (\ref{riccati-spinor}), respectively.

\section{Numerical Results}

\subsection{Results for zero carrier phase: $\varphi=0$.}
In this subsection, using the Riccati equation, we compute the momentum spectrum of the produced particles for several values of the chirp parameter with the carrier phase $\varphi$ set to $0$. (In all the subsequent computations we take $k_{\perp}$ to be zero.)

\begin{figure}[htb]
\begin{center}$
\begin{array}{cc}
\includegraphics[scale=0.6]{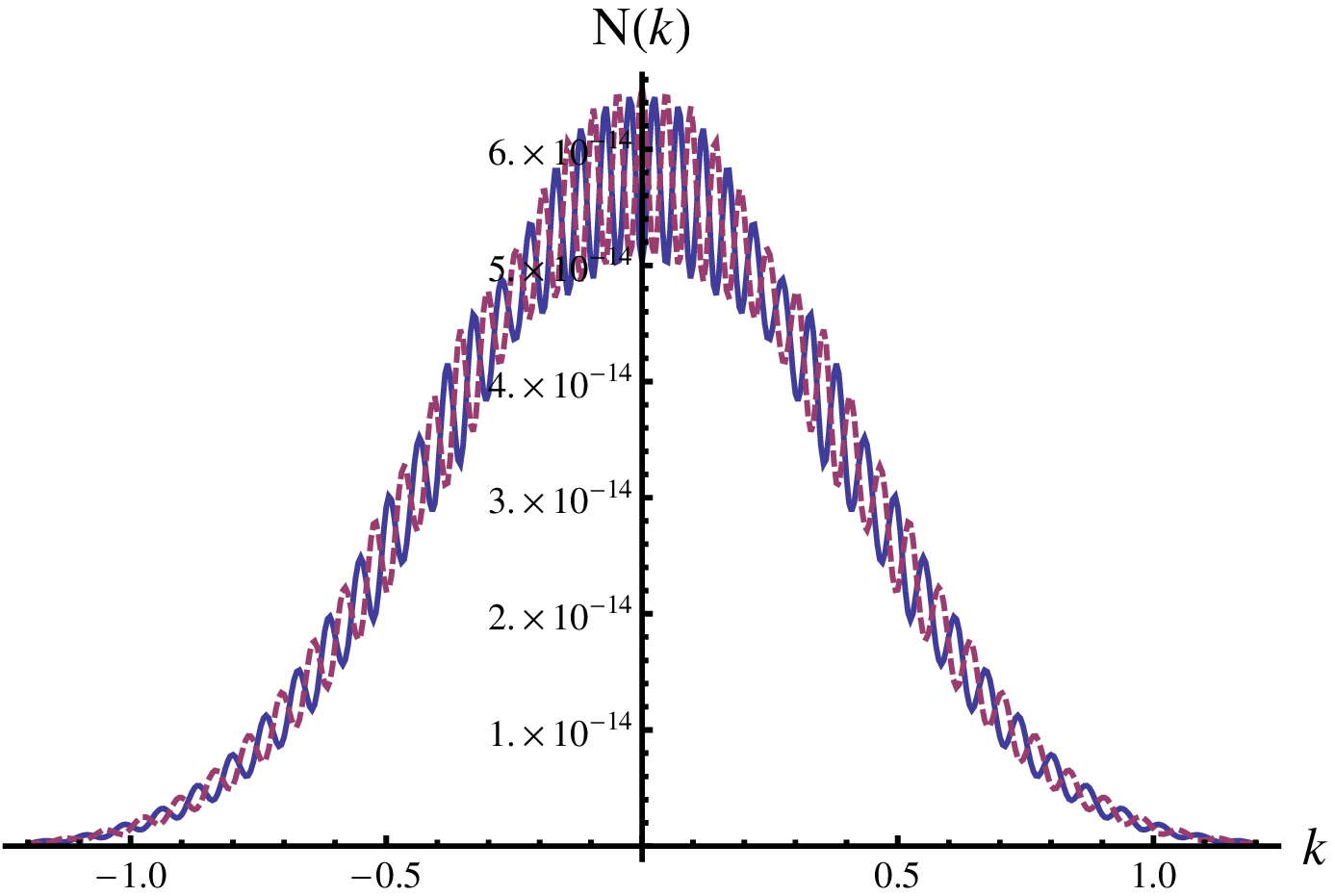}&
\includegraphics[scale=0.6]{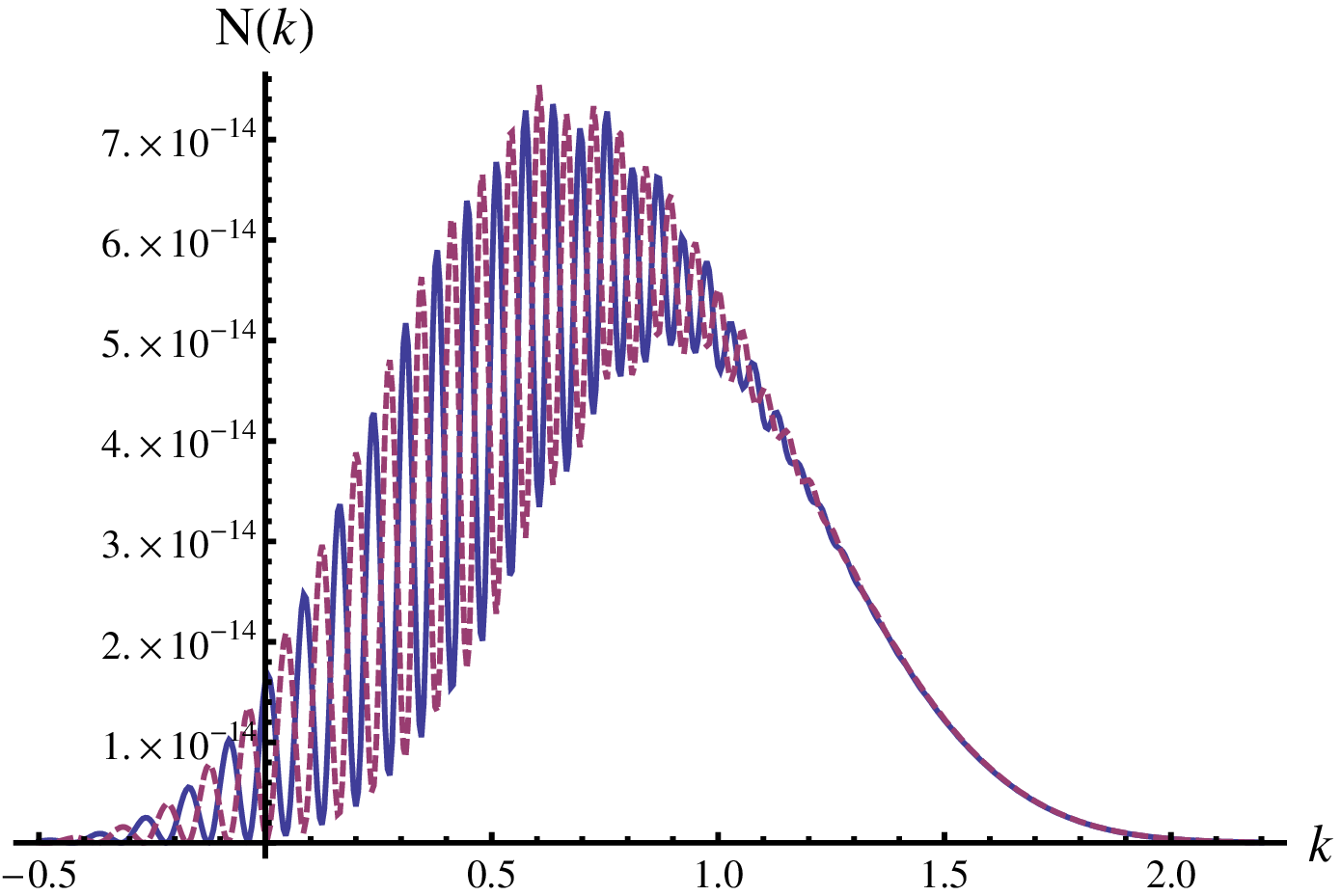}\\
\includegraphics[scale=0.6]{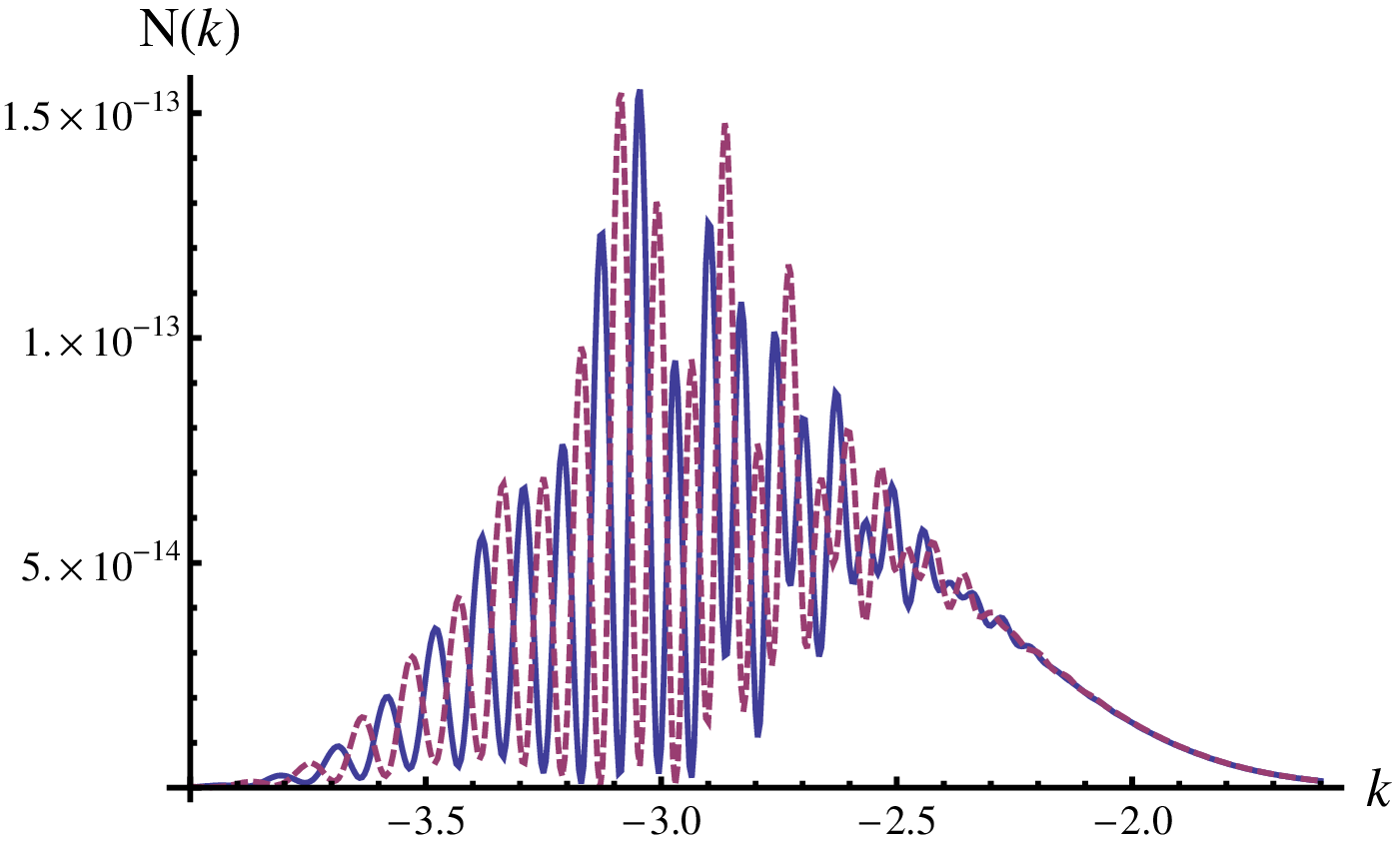}&
\includegraphics[scale=0.6]{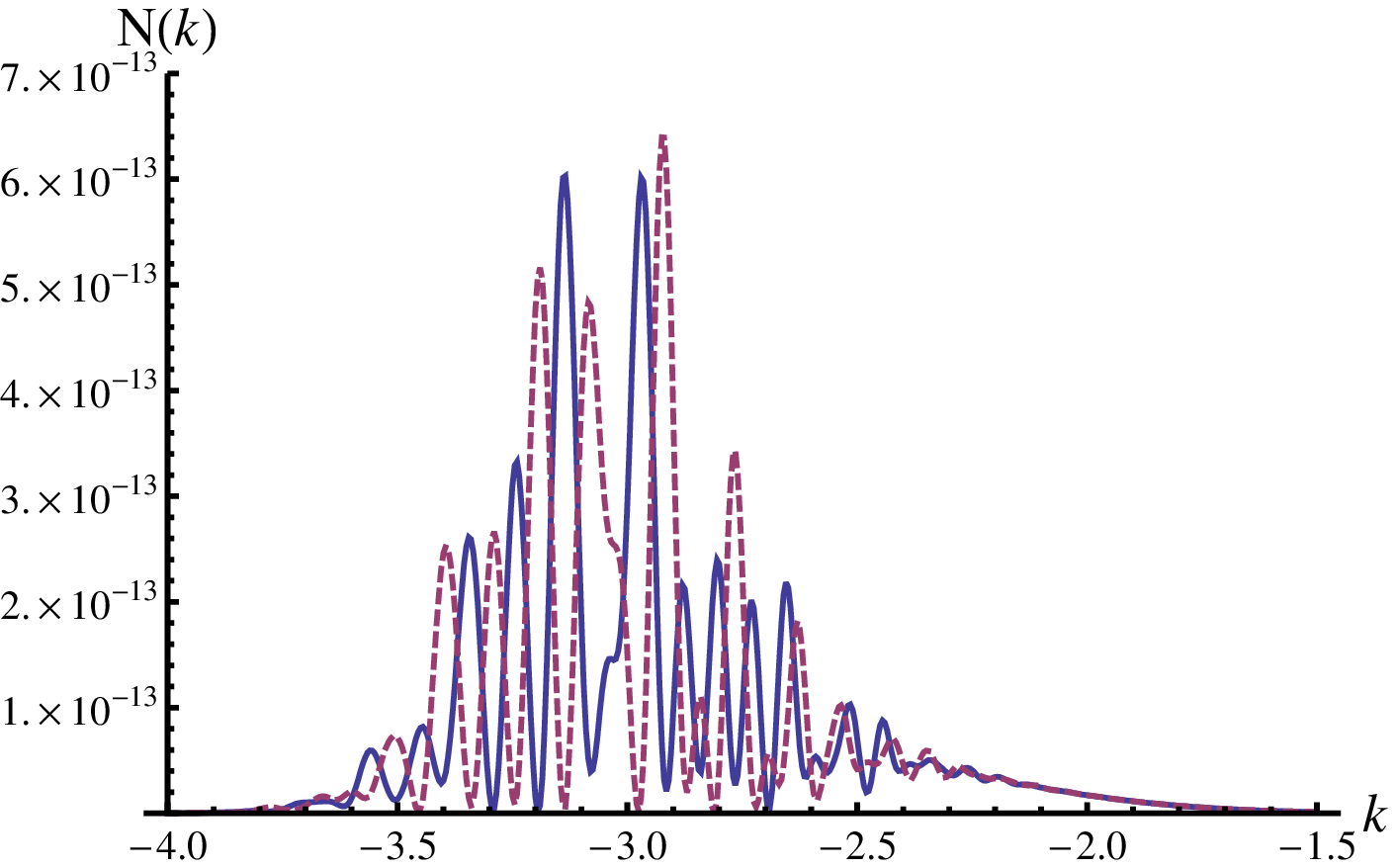}
\end{array}$
\end{center}
\caption{The thick [blue] and dashed [red] curves show the longitudinal momentum spectrum of the produced particles for the scalar and spinor cases, respectively. These plots are for zero carrier phase: $\varphi=0$. The other field parameters are $E_{0}=0.1$, $\omega=0.05$, and $\tau=100$, and from top left to bottom right the values of the chirp parameter are $b=0$, $b=0.00025$, $b=0.0005$, and $b=0.00075$. Note that, as in the carrier phase fields studied in \cite{Hebenstreit:2009km}, the spectra for spinor and scalar QED are out of phase. Also notice the strong sensitivity of the spectra to the value of the chirp parameter $b$.}
\label{phi0}
\end{figure}

For $b=0$, the momentum spectrum is centered around origin and oscillations are observed (see Fig. \ref{phi0} top left). The physical origin of the oscillations can be understood as an interference effect between temporally separated reflected waves due to multiple bump structure of the scattering potential. The WKB analysis of these effects is explained in \cite{cg} in terms of interference between separate complex conjugate pairs of turning points. When the chirp parameter $b$ set to $2.5\times10^{-4}$, the momentum spectrum gets shifted in the positive direction and the oscillations become more pronounced around bell shaped distribution (see Fig. \ref{phi0} top right). This result is similar to the effect introduced by the carrier phase in \cite{Hebenstreit:2009km}.  For the higher values of b the spectrum gets shifted in negative direction. The rate gets higher compared to the previous cases due to increase in the field's effective frequency. All the spectrums have the scalar and spinor results are off-phase due to double valued-ness of the spinor wave function \cite{cg}.

\begin{figure}[htb]
\includegraphics[scale=0.5]{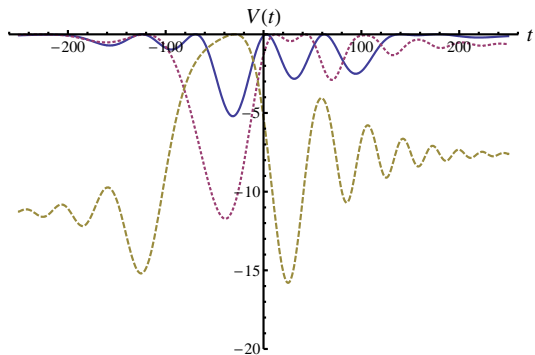}
\caption{The thick line shows the scattering potential with $b=0$ The other values of $b$ are chosen as $0.00025$(dotted) and $0.0005$(dashed). The longitudinal momentum is conveniently chosen as $p_{\parallel}=0.3$ to show the effect of chirp parameter on the bump structure of the potential. The other parameters are given as $E_{0}=0.1$, $\omega=0.05$, $\tau=100$, $\varphi=0$ (upper) and $\varphi=\pi/2$ (lower)}
\label{potential}
\end{figure}

These changes in the momentum spectrum are very sensitive to the value of the chirp parameter. The physical explanation for this in the scattering picture is that the corresponding effective scattering potential
\begin{equation}
V_k(t)=-\left(k-\frac{q}{c}A(t)\right)^{2},
\end{equation}
changes substantially even by small variations of the chirp parameter (see Fig. \ref{potential})

\subsection{Results for non-zero carrier phase: $\varphi=\frac{\pi}{2}$.}

\begin{figure}[htb]
\begin{center}$
\begin{array}{cc}
\includegraphics[scale=0.6]{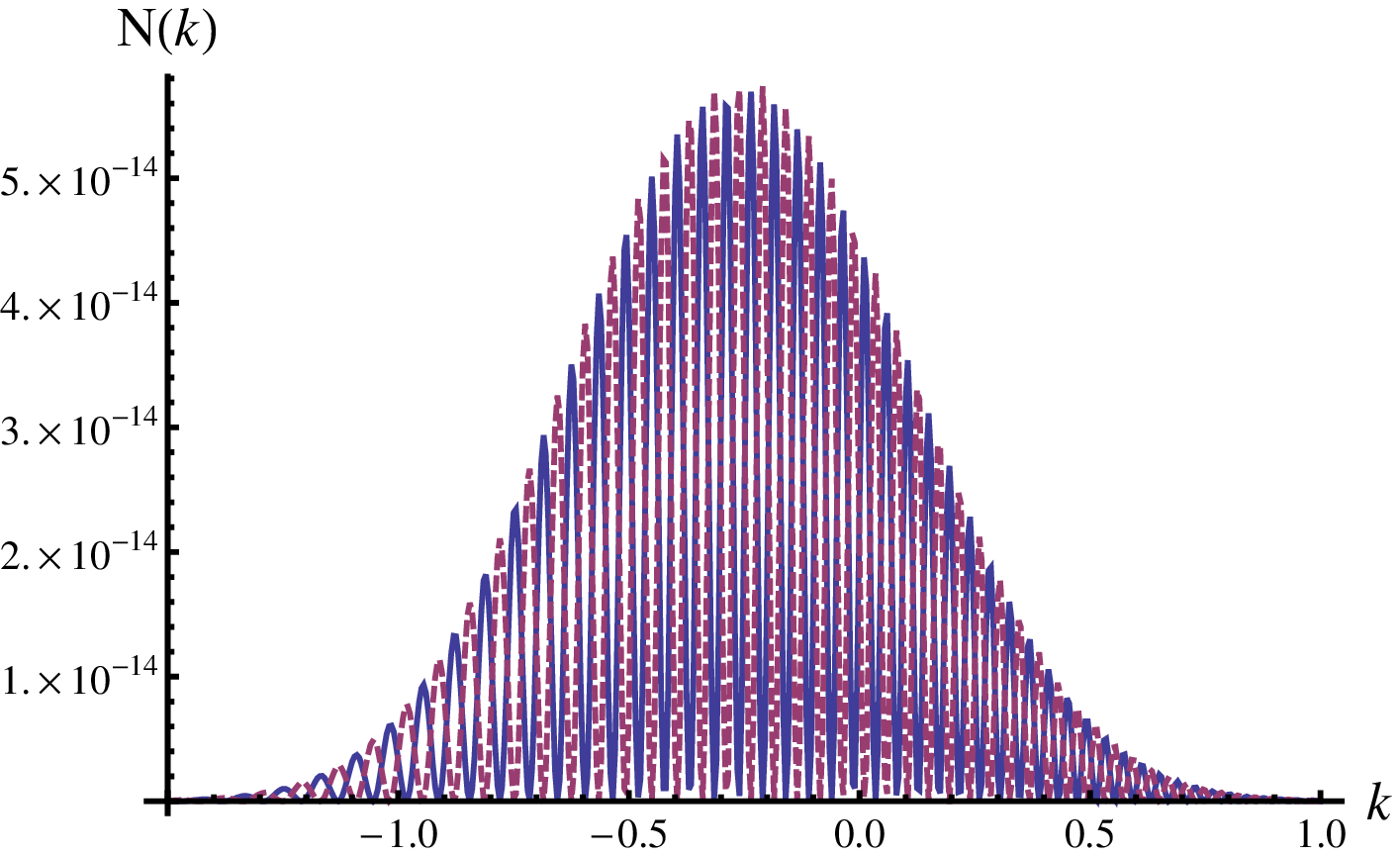}&
\includegraphics[scale=0.6]{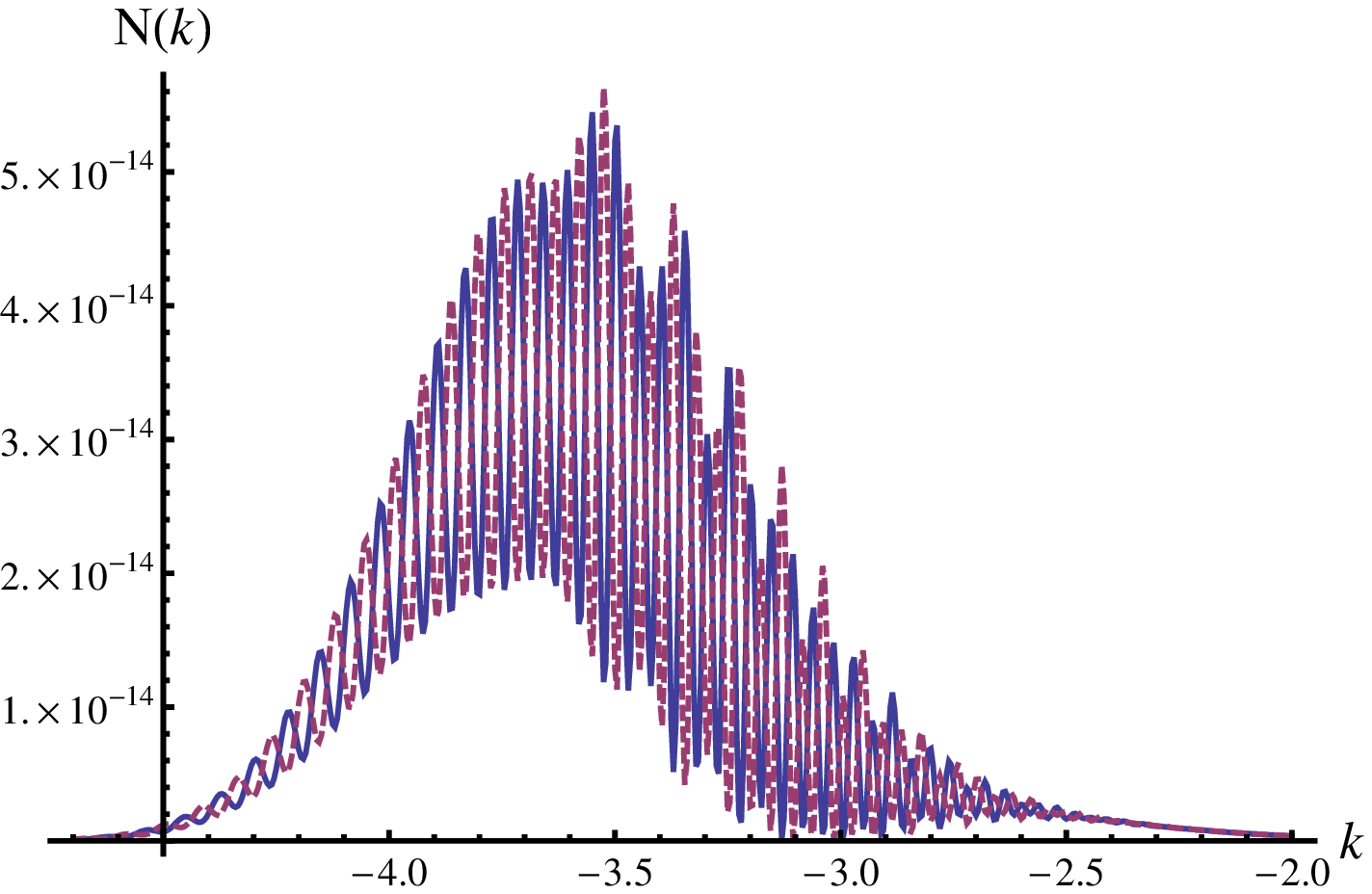}\\
\includegraphics[scale=0.6]{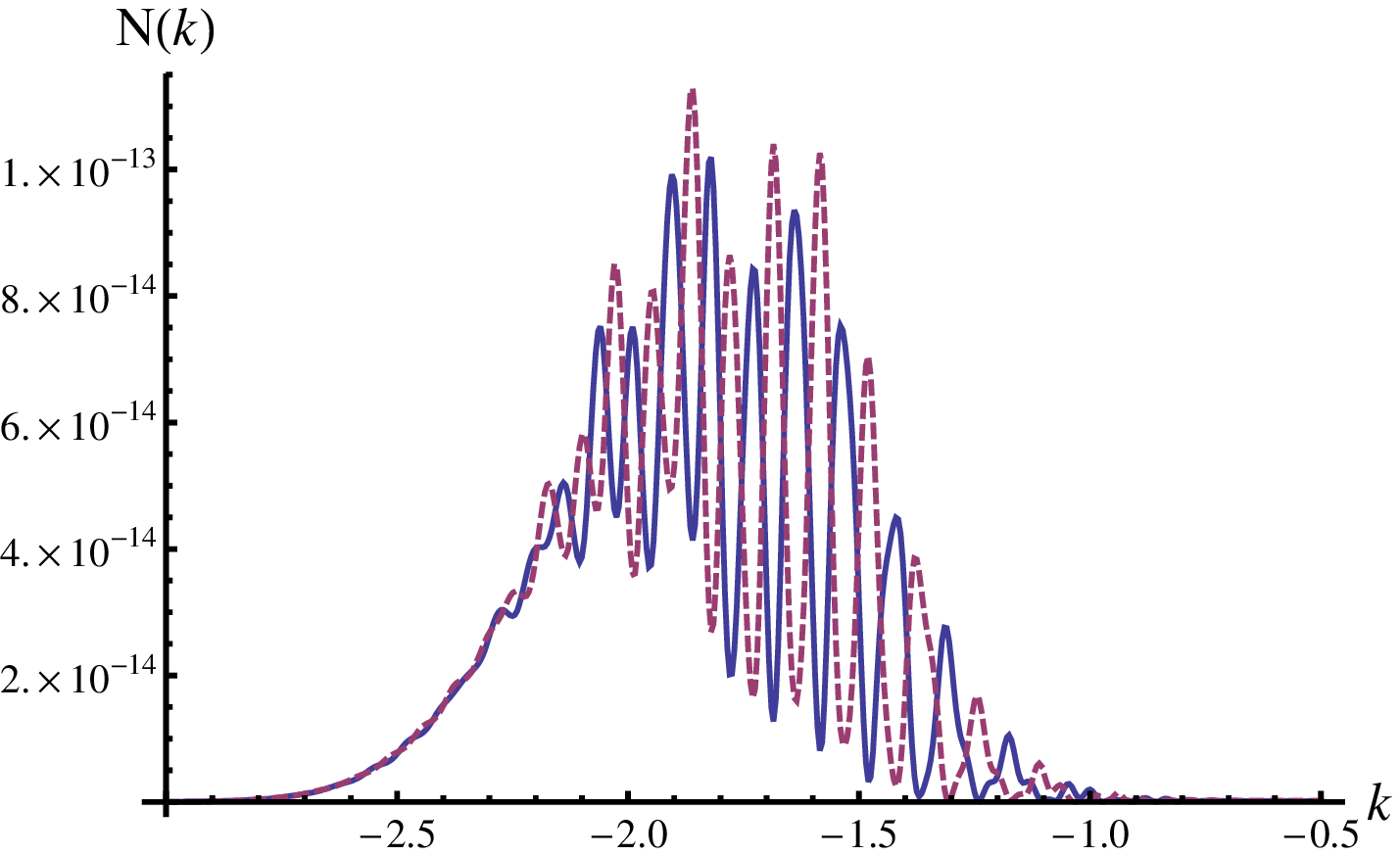}&
\includegraphics[scale=0.6]{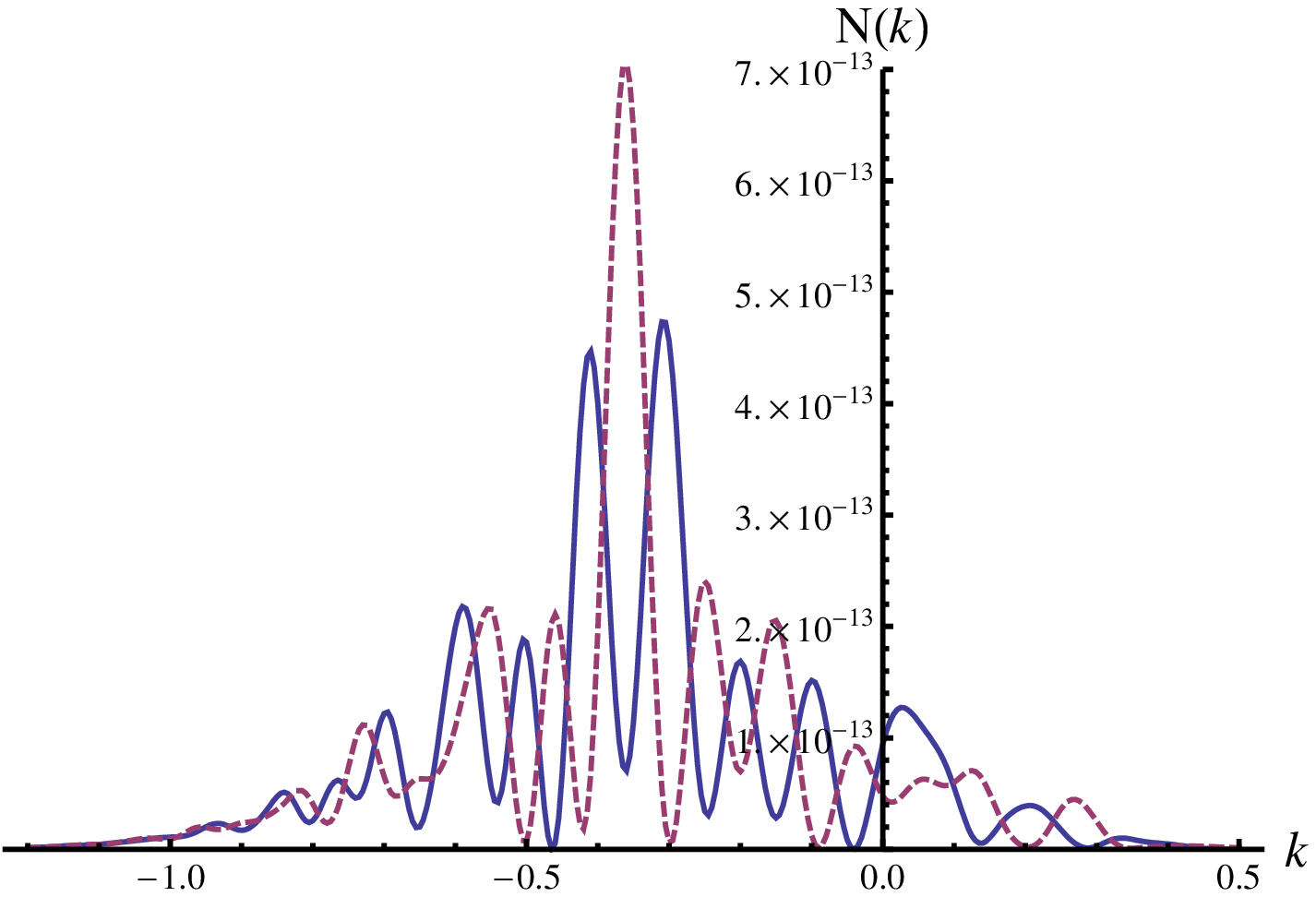}
\end{array}$
\end{center}
\caption{The thick [blue] and dashed [red] curves show the longitudinal momentum spectrum of the produced particles for the scalar and spinor cases, respectively. These plots are for non-zero carrier phase: $\varphi=\pi/2$.  The other field parameters are $E_{0}=0.1$, $\omega=0.05$, and $\tau=100$, and from top left to bottom right the values of the chirp parameter are $b=0$, $b=0.00025$, $b=0.0005$, and $b=0.00075$. Note that, as in the carrier phase fields studied in \cite{Hebenstreit:2009km}, the spectra for spinor and scalar QED are out of phase. Also notice the strong sensitivity of the spectra to the value of the chirp parameter $b$.}
\label{phipi2}
\end{figure}

To see the effect of carrier phase accompanied with the chirp, we set $\varphi$ to $\pi/2$ and use the previous values for the chirp parameter. When the chirp parameter is set to $0$, the scattering barrier becomes completely transparent \cite{Hebenstreit:2009km} for various values of momenta which results in almost zero reflection probability (see Fig. \ref{phipi2} top left). When $b$ is set to $2.5\times10^{-4}$ the momentum spectrum is shifted in negative direction (see Fig. \ref{phipi2} top right). The oscillatory structure of the momentum spectrum resembles that of with the parameters $b=0$ and $\varphi=\pi/4$ \cite{Hebenstreit:2009km}. Considering this fact and the result represented in Fig. \ref{phi0} (top right), for relatively small values of $b$, chirp parameter effectively plays the role of carrier phase. However for larger values of $b$, the oscillations does not have a well defined envelope; there appears to be two main peaks in the distribution for scalar case and a single peak for spinor case when b is set to $7.5\times10^{-4}$ (see Fig. \ref{phipi2} bottom right).

\subsection{Turning-point structure}

In this section application of complex WKB to vacuum pair production will be discussed briefly. The results of simple gauge fields will be extended by analogy to give a qualitative explanation of the computed momentum spectrum of the produced particles \cite{cg}.

As discussed in Section I.B, each mode of the field operator satisfies a Schr\"odinger-like (\ref{schro}) equation and in the scattering picture, the particle number is related to the reflection coefficient, as in (\ref{riccati-scalar}) and (\ref{riccati-spinor}).  Therefore, in this approach, complex WKB method \cite{heading,Berry:1972na,meyer}. can be used as an analytic way of obtaining the particle number for the specified momentum. In obtaining the reflection coefficient, WKB solutions are traced from $t\rightarrow-\infty$ to $t\rightarrow\infty$ to find the asymptotic ratio of the WKB coefficients. The dominant contribution to coefficients is obtained when the wave function is traced around the points where effective scattering potential $\omega(t)$ is zero. These points are called turning points and they appear in complex conjugate pairs when the potential is real. The simplest case assumes that the scattering potential just has turning points located on the imaginary axis in complex conjugate pairs. Previously studied  gauge potential satisfying this condition is given as \cite{popov,Dumlu:2009rr,kimpage}:
\begin{equation}
A(t)=E_{0}\tau\tanh\left(\frac{t}{\tau}\right)
\end{equation}
where $\tau$ is the pulse width. WKB result for the particle number density is given as \cite{heading,landau}:
\begin{equation}
N_{\rm}\approx e^{-2K} , \qquad K=\left| \int_{t_1}^{t_2} \omega(t)\, dt\, \right |,
\end{equation}
where $t_1$ and $t_2$ are the dominant (ie. closest to the real axis) turning points of the corresponding scattering potential. The location of the turning points can be viewed as a function of longitudinal momentum $k$ with $\tau$ kept fixed, and the pair production rate is given as a function of momentum. The momentum spectrum is an exponentially decaying distribution and it is in good agreement with the exact result in the quasiclassical regime \cite{Dumlu:2009rr}. A more general model that exhibits the resonance effect has four turning ($t_1$..$t_4$) points where the gauge potential is given as \cite{cg}:
\begin{equation}
A(t)=\frac{E_{0}}{\tau(1+\frac{t^2}{ \tau^2})}.
\end{equation}
Here $E_0$ is the intensity and $\tau$ represents the width of the field. Pair production rate has been found as \cite{cg,froman}:
\begin{equation}
N_{\rm}\approx e^{-2K_1}+e^{-2 K_2}\pm 2\cos(2\alpha)\,e^{-K_1-K_2}
\label{rate}
\end{equation}
where $+$ and $-$ signs represent the scalar and spinor cases respectively. The other quantities are defined as:
\begin{eqnarray}
K_1&=&\left| \int_{t_1}^{t_2} \omega(t)\, dt\, \right | \quad,\quad
K_2=\left| \int_{t_3}^{t_4} \omega(t)\, dt\, \right | \nonumber\\
\alpha&=& L-\sigma(K_1) -\sigma(K_2) \nonumber \\
L &=&\left| {\mathcal Re}\left( \int_{t_2}^{t_3} \omega(t)\, dt\right) \right | \nonumber\\
\sigma(K)&=&\frac{1}{2}\left[\frac{K}{\pi}\left(\ln \left(\frac{K}{\pi}\right)-1\right)+{\rm Arg}\,\Gamma\left(\frac{1}{2}-i\frac{K}{\pi}\right)\right]\nonumber
\end{eqnarray}
where the limits of integration correspond to turning points.  The result above, involves exponentially decaying terms accompanied by an interference term connecting the neighboring turning points (see Eqn. \ref{rate}). The interference term is responsible for the oscillations observed in the momentum spectrum.

The distribution of the turning points can provide qualitative information about the momentum spectrum. In general, the phase integral $K$  is almost linear in $t$ therefore, the dominant contribution to reflection coefficient comes from the terms involving turning points which are closest to the real axis. In this model, all the turning points are at the same distance to real axis therefore, interference term in (\ref{rate}) is of the same order of magnitude with the the other terms. This fact explains why for several values of momenta there can be minima which are close to zero in the momentum spectrum \cite{cg}.

\begin{figure}[htb]
\begin{center}$
\begin{array}{cc}
\includegraphics[scale=0.75]{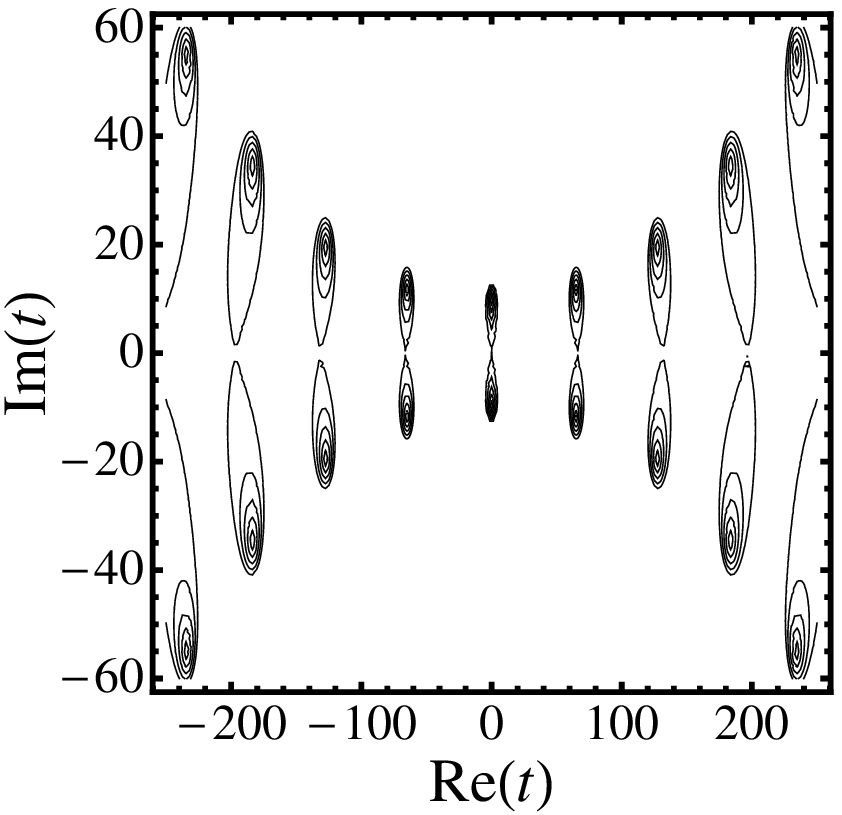}&
\includegraphics[scale=0.75]{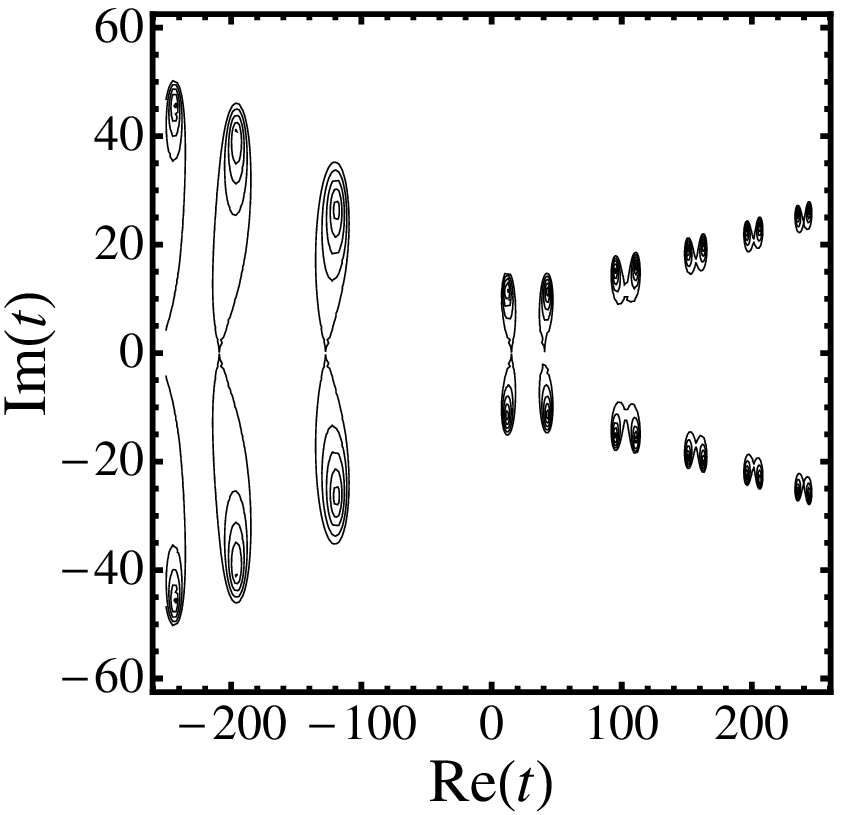}\\
\includegraphics[scale=0.75]{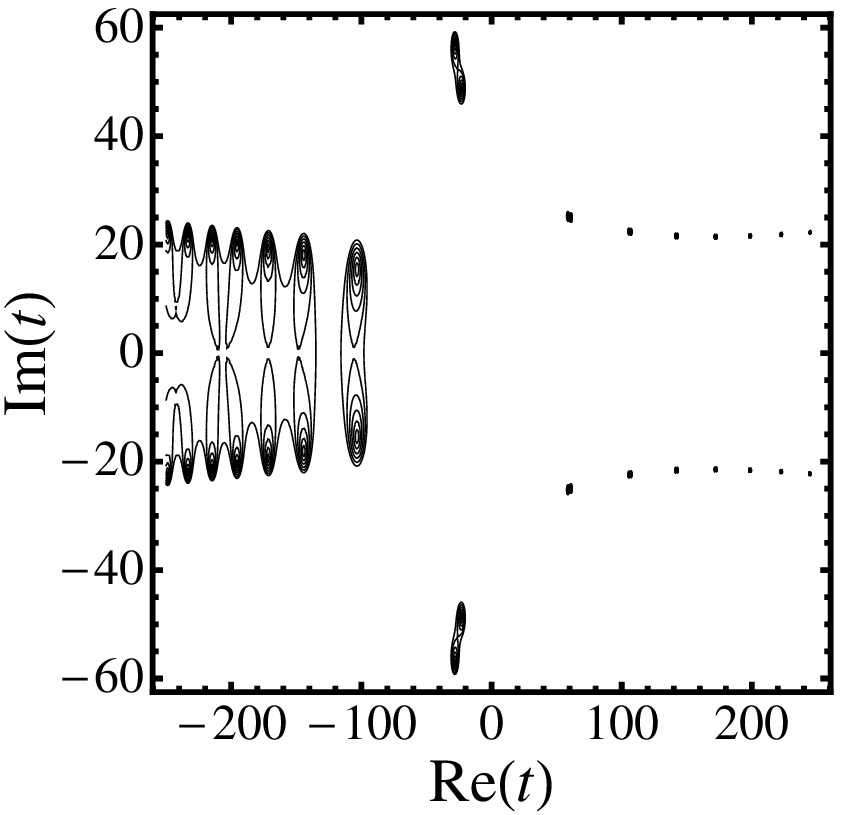}&
\includegraphics[scale=0.75]{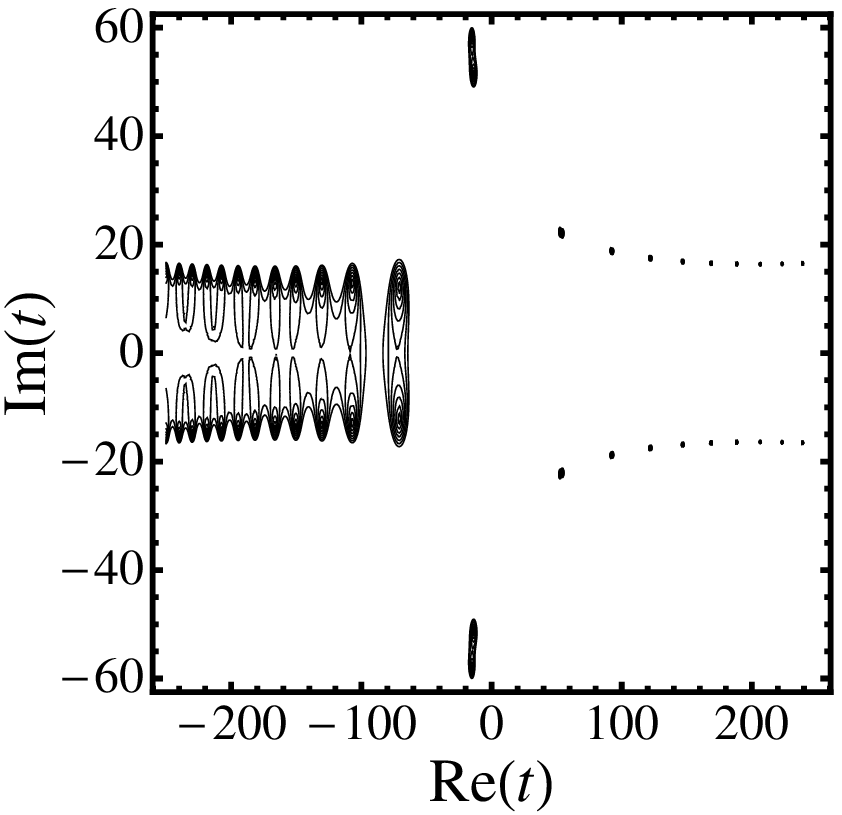}
\end{array}$
\end{center}
\caption{Contour plots of $|\omega_{\bf k}(t)|^2$ in the complex $t$ plane, showing the location of turning points where $\omega_{\bf k}(t)=0$. These plots are for zero carrier phase: $\varphi=0$.  The other field parameters are $E_{0}=0.1$, $\omega=0.05$, and $\tau=100$, and from top left to bottom right the values of the chirp parameter are $b=0$, $b=0.00025$, $b=0.0005$, and $b=0.00075$, and longitudinal momentum values $k=0$, $k=0.5$, $k=-3.0$ and $k=-3.0$, respectively. Notice the strong sensitivity of the locations of the turning points to the value of the chirp parameter $b$.}
\label{phi0tp}
\end{figure}

\begin{figure}[htb]
\begin{center}$
\begin{array}{cc}
\includegraphics[scale=0.75]{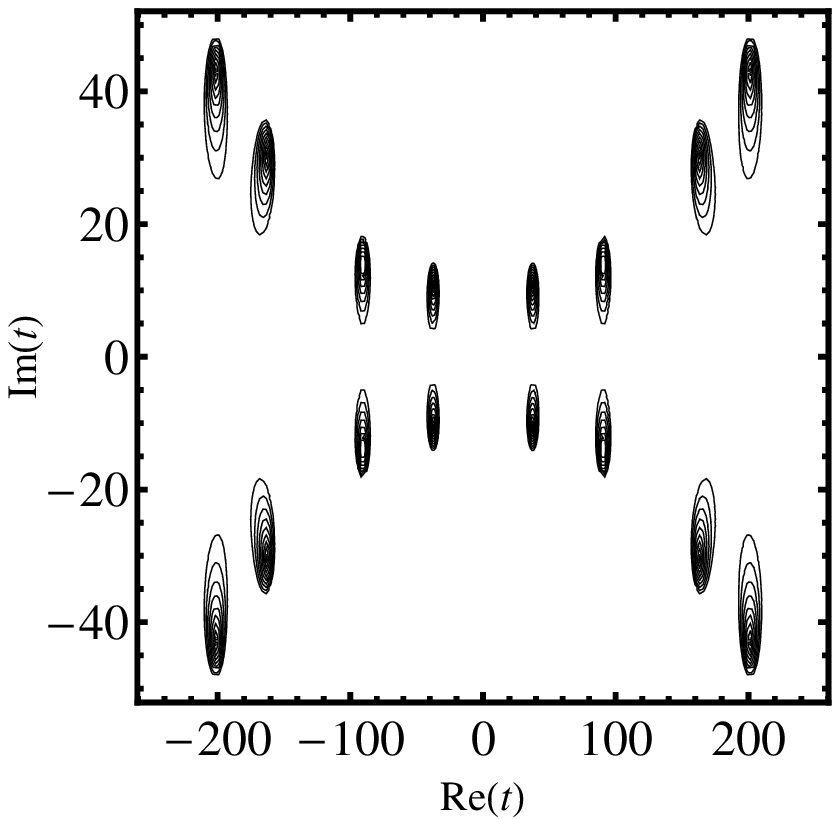}&
\includegraphics[scale=0.75]{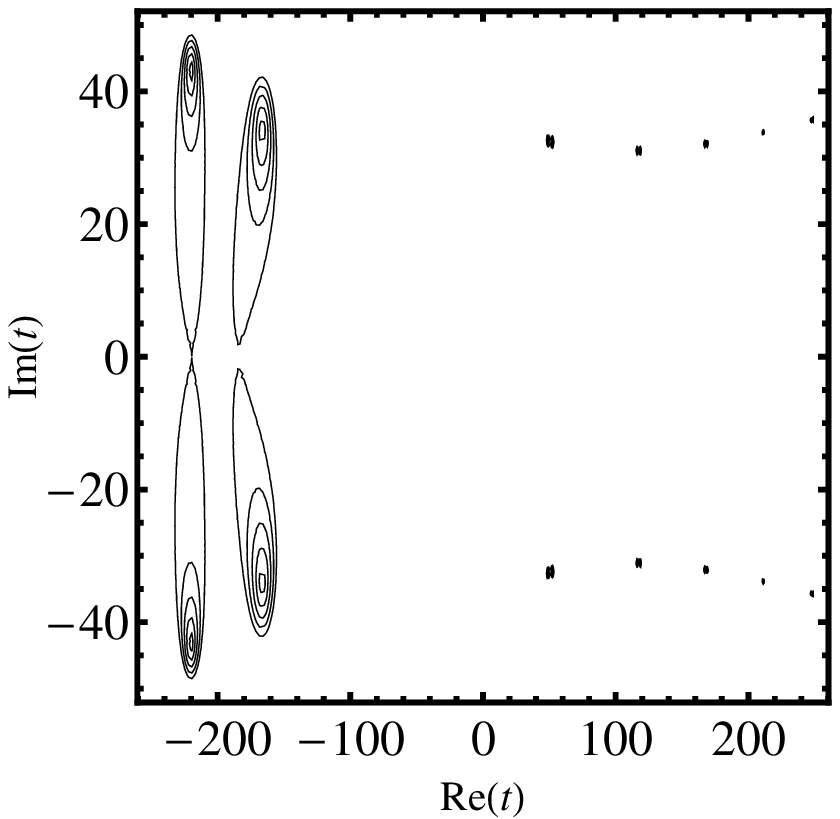}\\
\includegraphics[scale=0.75]{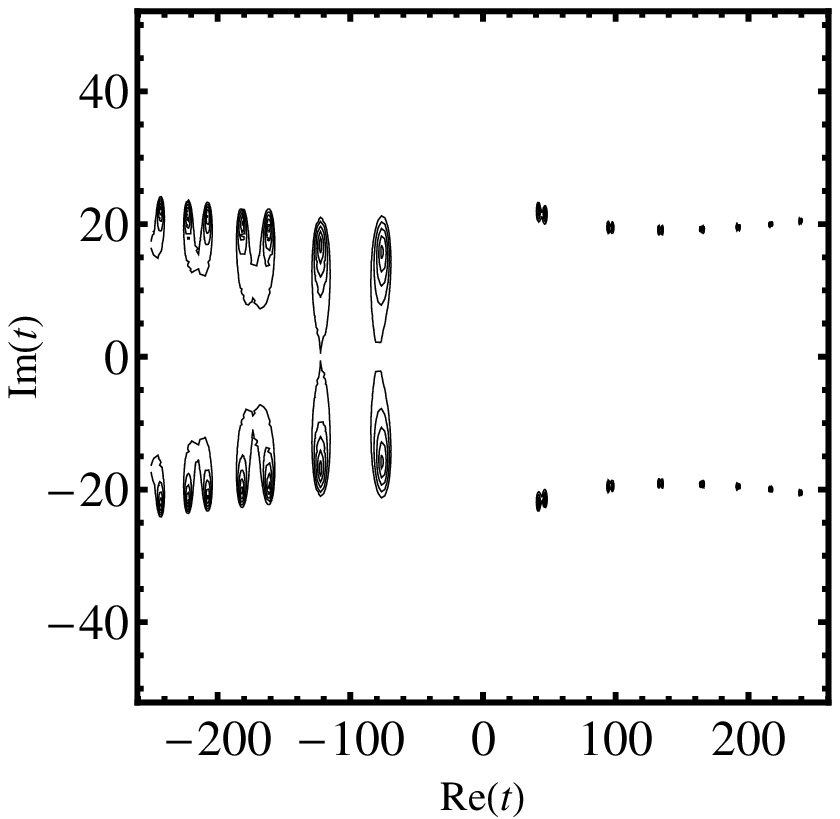}&
\includegraphics[scale=0.75]{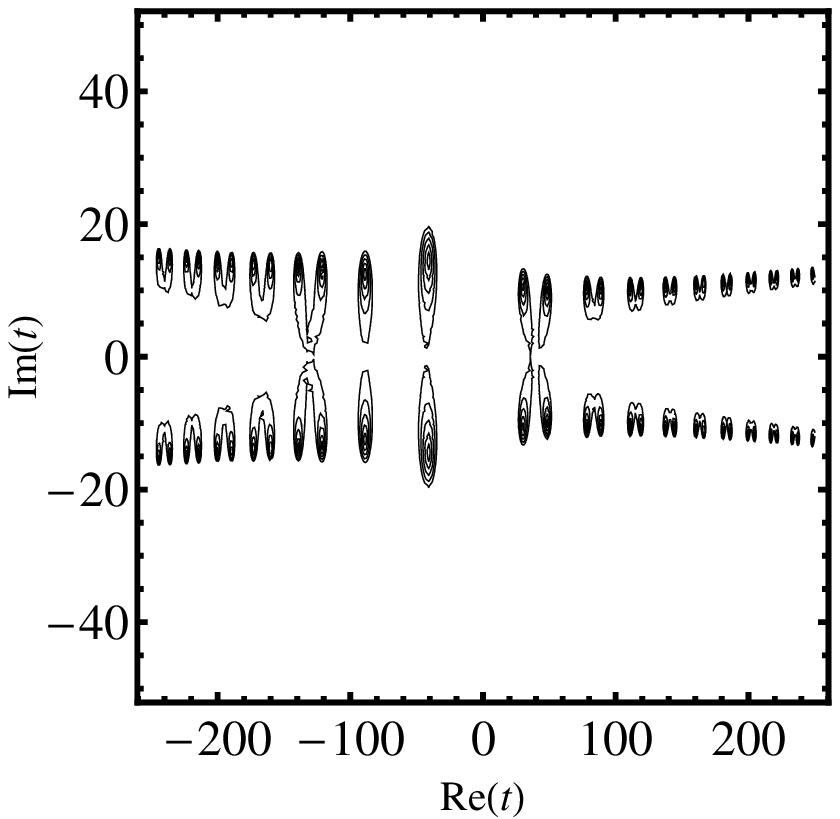}
\end{array}$
\end{center}
\caption{Contour plots of $|\omega_{\bf k}(t)|^2$ in the complex $t$ plane, showing the location of turning points where $\omega_{\bf k}(t)=0$. These plots are for non-zero carrier phase: $\varphi=\pi/2$.  The other field parameters are $E_{0}=0.1$, $\omega=0.05$, and $\tau=100$, and from top left to bottom right the values of the chirp parameter are $b=0$, $b=0.00025$, $b=0.0005$, and $b=0.00075$, and longitudinal momentum values $k=-0.5$, $k=-3.5$, $k=-2.0$ and $k=-0.5$, respectively. Notice the strong sensitivity of the locations of the turning points to the value of the chirp parameter $b$.}
\label{phipi2tp}
\end{figure}

These conclusions can be extended to any field configuration without loss of generality. As for the chirped field, there are infinite number of turning points and the relevant ones are shown in Fig. \ref{phi0tp} and \ref{phipi2tp} for various chirp and momentum values. For $\varphi=0$ and $b=0$, it is seen that the dominant contribution comes from single central pair of turning points. The magnitude of the resonances in the corresponding momentum spectrum (see Fig. \ref{phi0} top left) are small compared to the overall magnitude of the pair production rate. The reason is that interference terms are suppressed by the exponential terms and these terms involve turning points which are further away from the real axis than the central pair of turning points (see Fig. \ref{phi0tp} top left). Therefore the magnitude of the interference terms are small compared to contribution coming from the central pair. On the other hand, for $\varphi=\pi/2$ and $b=0$, the dominant contribution comes from the two central pairs of turning points (Fig. \ref{phipi2tp} top left). They are equally far away from the real axis as in the model discussed and the argument for four turning point case basically explains why there are minima close to zero in the corresponding momentum spectrum (see Fig. \ref{phipi2} top left).

The effect of the chirp parameter on the spectrum can also be explained by analyzing the distribution of the turning points as shown In Fig. \ref{phi0tp} and \ref{phipi2tp}. As the chirp parameter changes, the momentum values for which dominating pair of turning points gets closest to real axis also change. This results in a shift of the spectrum along the momentum axis. The distribution of turning points for non-zero chirp seems to be more complicated; the dominant contribution comes from multiple pairs of turning points with multiple interference terms being effective on the momentum spectrum. This results in a complicated oscillatory structure.

\section{Conclusions}

In conclusion, the two main effects of the chirp parameter on the momentum spectrum can be summarized as shifting the spectrum along the momentum axis and increasing the magnitude of the number density at certain momenta. The chirp parameter, for relatively small values, has a similar effect on the spectrum as the carrier phase. However, when the chirp is increased, the oscillatory behavior of the spectrum gets more complicated. The basic features of the computed momentum spectrum such as the shift and the oscillatory structure can qualitatively be understood by analyzing the distribution of the turning points. The results by analogy can be extended to realistic field configurations. This enables to understand the effect of the important field parameters such as the carrier phase and the chirp parameter thereby giving useful insight into the structure of external pulse.
\\

This study was supported by DOE grant DE-FG02-92ER40716. The author thanks G. Dunne for fruitful discussions.

\end{document}